\documentstyle[12pt]{article}

\begin{document}

\newcommand{\lr}{\longrightarrow}
\newcommand{\ra}{\rightarrow}
\newcommand{\cg}{{\cal G}}
\newcommand{\ol}{\overline}
\newcommand{\bfm}{{\bf m}}
\newcommand{\bfn}{{\bf n}}
\newcommand{\ds}{\displaystyle}
\newcommand{\cu}{{\cal U}}
\newcommand{\bone}{{\bf 1}}
\newcommand{\botwo}{{\bf 2}}
\newcommand{\bothree}{{\bf 3}}
\newcommand{\boeign}{{\bf 8N}}

\title{Family Unification from Universality}
\author{P.P. Divakaran \\
Chennai Mathematical Institute \\
92 G.N. Chetty Road, T. Nagar \\
Chennai-600 017, India. \\
E-mail: ppd@smi.ernet.in}
\date{}
\maketitle

\begin{center}
{\Large \bf Abstract}
\end{center}

\vspace{2mm}

A direct consequence of the occurrence of fermion families in
the standard model is the invariance of fermion currents
under certain groups of (universality) transformations.  In
this paper we show how these universality properties can
themselves be used as a method of finding and studying
``grand'' family unification models.  In the exact standard
model limit two independent universality groups $S_{wk}$ and
$S_{st}$ of weak and strong gauge interactions are first
identified.  The subgroup of any family unification group
$\cg$ whose currents are invariant under $S_{wk} (S_{st})$
is then the centraliser $G_{wk} (G_{st})$ of $S_{wk}
(S_{st})$ in $\cg$.  Choosing $\cg = SU(8N)$, we find
$G_{wk} = SU(2)$ and $G_{st} = U(1) \times SU(3)$; 
the standard model group $G = G_{wk} \times G_{st}$ is the
group which respects either weak or strong universality.
The fundamental representation {\bf 8N} of $SU(8N)$
decomposes under $SU(2) \times SU(3)$ as $N$ copies of $({\bf
2}, {\bf 1}) \oplus (\botwo,\bothree )$; their $U(1)$ charges are the
usual hypercharges.  A Higgs field transforming as $\boeign$
accomplishes the secondary symmetry reduction (from $G$ to
$U(1)_{em} \times SU(3))$ satisfactorily.  The requirement
that charged currents be $V-A$ forces all fermions to be
left-handed in the unbroken $G$ limit, making the model
chirally invariant in a strong sense -- fermions have no
right-handed components to make masses with.  The remedy
proposed is that $R$-fermions are composites of $L$-fermions
and Higgs.  If their binding is in one specific channel,
$R$-fermions are shown to come in the right numbers and with
the right couplings to ensure pure $V$ $U(1)_{em}$ and $SU(3)$
currents while leaving the $SU(2)$ currents unchanged.  It
is finally argued that universality is most naturally
understood in terms of a simple preonic structure for
fermions (but not for gauge bosons), obviating the need for
a primary $\cg \ra G$ Higgs mechanism.  $SU(8N)$ is then
best interpreted as the global ``metaflavour'' group of
$L$-chiral fermions.  In this picture, $\cg /G$ is not gauged;
there are no ultraheavy gauge bosons and hence no anomaly or
hierarchy problem.

\section{Introduction}

The problem of accommodating the existence of several
replicas of a family of leptons and quarks in unified theories
of all their interactions (except, of course, gravitational)
has preoccupied model makers for more than two decades now.
Beginning with {\it ad hoc} impositions of invariance under
(finite [1] or Lie [2]) groups of so called horizontal
symmetries, this endeavour soon moved on to the recognition
that certain orthogonal groups $\cg$ have subgroups $G$ and
representations $\rho$ such that the restriction of $\rho$
to $G$ is a direct sum of copies of a unique irreducible
representation of $G$, making such groups plausible candidates
for family unification [3].  Such models, however, have
certain fundamental difficulties (primarily to do with
unacceptably large right chiral interactions) which have
resisted a satisfactory natural resolution.  Subsequently,
attention turned increasingly to supersymmetric models; in
particular, very detailed work on superstring - inspired
family unification models are going on apace now [4].  At
the same time it has recently been shown that conventional
nonsupersymmetric unification models, when combined with the
topological properties of the Higgs phase of nonabelian gauge
theories -- i.e., going beyond the perturbative regime -- have
the richness to accommodate the family structure [5,6].

The approach of the present paper to dealing with families
is, by and large, orthodox, at least in its fundamentals;
supersymmetry is not invoked and topological aspects of the
reduction of a unifying gauge group $\cg$ to an observed or
effective gauge group $G$ are ignored in the Lie algebraic
considerations below.  It is also a pragmatic and
phenomenological approach.  Instead of looking for the source
of the family structure, we shall focus primarily on its most
characteristic empirical signature, namely the observed
universality properties of the gauge interactions of all
leptons and quarks, $8N$ in number where $N$ is the number
of families.  We shall seek to determine a simple group
$\cg$ such that the subgroup of $\cg$ whose currents respect
the demand of universality is the (unbroken) standard model
group $G = SU(2) \times U(1) \times SU(3)$.  In the actual
implementation, as described in the next section, we proceed
as follows.  Ignoring fermion masses and mixings, i.e.,
before $G$ is further broken to $U(1)_{em}\times SU(3)$, the
$G$-gauge interactions of fermions are separately invariant
under two types of -- weak and strong -- universality
transformations forming two distinct subgroups $S_{wk}$ and
$S_{st}$ of $\cg$.  The $\cg$-gauge currents which are
invariant under $S_{wk}$ (respectively $S_{st}$) are easily
shown to couple to the gauge bosons of a subgroup $G_{wk}
(G_{st})$ of $\cg$ which is the centraliser of
$S_{wk}(S_{st})$, namely the group of elements of $\cg$ which
commute with all elements of $S_{wk}(S_{st})$.  Moreover,
the gauge group whose currents are invariant under {\it
either} $S_{wk}$ {\it or} $S_{st}$ is $G_{wk} \times
G_{st}$.  After identifying $S_{wk}$ and $S_{st}$ (taking
account in particular of the fact that leptons have no
strong interactions), it is established that if $\cg$ is
chosen to be $SU(8N)$, then $G_{wk} = SU(2)$ and $G_{st} =
U(1) \times SU(3)$.  Conversely if we want to have $G =
G_{wk} \times G_{st}$ to be $SU(2) \times U(1) \times
SU(3)$,  then $\cg$ can only be $SU(8N)$.  The fundamental
representation $\boeign$ decomposes under $SU(2) \times SU(3)$ as
$N$ copies of the representation $(\botwo,\bone ) \oplus
(\botwo,\bothree)$ i.e, $N$ families of leptons and quarks.  Thus,
by embedding $SU(2) \times U(1) \times SU(3)$ in $SU(8N)$
{\it via} the imposition of weak and strong universalities,
$SU(8N)$ is made to play the role of a ``grand'' family
unification group (no other simple group can serve this
purpose) [7].  In particular, independently of the mechanism
for the reduction of the gauge group from $\cg$ to $G$ (more
fundamentally, even in ignorance of the true source of
universality), the relative coupling constants of $SU(2)$,
$SU(3)$ and $U(1)$ get fixed.

It is noteworthy that the (``weak'') hypercharge $U(1)$
reflects the absence of strong interactions for the leptons
(as it perhaps should since, empirically, it does
distinguish between leptons and quarks).  Other indications
that the model is somewhat off the beaten track also become
apparent at this point.  For instance, the $SU(2)$ coupling
is smaller, in relation to $SU(3)$ and $U(1)$, by a factor
$\sqrt{2}$ than in most conventional unification models,
resulting in a (unification) value of $\sin^2\theta$ of
3/4.  Moreover, since $SU(2)$ currents must be left-chiral,
all fermions are forced to be left-handed (on the scale of
the $G$ gauge theory), making the $U(1)$ and $SU(3)$ currents
also left-chiral.  Thus, in the $G$-gauge invariant limit,
all fermions are massless and chiral symmetry is exact in
the strongest possible form: there are no right-handed
($R$-) fermions to couple to $L$-fermions 
so as to generate masses.  This circumstance suggests
strongly that the mechanisms responsible for the restoration
of parity invariance of electromagnetic and colour
interactions and for the appearance of fermion masses are
one and the same, namely a suitable dynamical breaking of
chiral symmetry, and that this mechanism is closely linked to
the spontaneous breaking of $SU(2) \times U(1) \times SU(3)$
to $U(1)_{em} \times SU(3)$.  

As a first step towards justifying such a linkage, we fall
back on the standard Higgs mechanism for breaking $SU(2)
\times U(1)$ to $U(1)_{em}$.  Accordingly, in the $\cg$
theory, we take the Higgs to constitute one $\boeign$
representation, decomposing as $N$ families of leptonic
Higgs $(\sim ( \botwo,\bone)$ under $SU(2) \times SU(3)$) and
$N$ of coloured Higgs $(\sim (\botwo ,\bothree))$.  On assigning
vacuum values to all $N$ leptonic Higgs in, say, the up
flavour, the model becomes indistinguishable from the
standard model except, of course, for the exact $(L$-)
chiral invariance. But once elementary scalars are admitted,
we have the possibility of generating $R$-fermions as
composites of $L$-fermions and Higgs [8] and, thence, of
producing masses by Yukawa couplings.  Details of the
dynamics of the proposal are beyond the essentially
group-theoretic scope of this paper.  It is quite easy to
show however that there is a ``channel' in the $L$-fermion
Higgs system which, if assumed attractive, can produce just
the right number of $R$-fermions with the right
transformation properties under $SU(2)$ and $SU(3)$: there
are $8N$ $R$-fermions, all of them transforming trivially
under $SU(2)$, breaking up as $2N$ multiplets each
transforming as $\bone \oplus \bothree$ under $SU(3)$.
Consequently, no $R$-fermion couples to $SU(2)$, the
$R$-leptons do not couple to $SU(3)$ while the $R$-quarks
do, with strength equal to that of $L$-quarks.  In other
words, $SU(2)$ currents are $V-A$ and the $SU(3)$ (quark)
currents pure $V$.  Finally, the $U(1)_{em}$ coupling
strengths of $R$-fermions are also easily computed; they
match precisely those of the corresponding $L$-fermions,
making the electromagnetic current also pure $V$.  

In the last section, we turn to the problem of the physics
underlying the reduction of the primitive group $\cg$ to the
standard model group $G$.  After exhibiting a simple Higgs
representation that will achieve this, a more directly
physical explanation for the manifestation of universality is
sought in the very natural idea that the $(L$-) fermions are
themselves composites of a set of preons or metafermions.  The simplest
explicit preonic model with built-in universality then
requires $6+N$ preons (2 for flavour, 4 for colour
including leptonic colour and $N$ for family) [9].  All of
them have meta-interactions binding them into massless,
$L$-chiral, leptons and quarks, but 6 of them have,
additionally, $SU(2) \times U(1) \times SU(3)$ gauge
interactions -- the photon, gluons, $W$ and $Z$ are thus
elementary.  In this perspective, our unifying $SU(8N)$ is
just 't Hooft's metaflavour group [10] unbroken as
long as fermions are massless, but broken softly by the
spontaneous generation of $R$-fermions and the concomitant
masses -- the gauging of only a subgroup of the metaflavour
group is no more than a reflection of the fact that some of
the preons already have such gauge interactions.

Obviously, these speculative ideas, touched upon only briefly in
the concluding section, need to be worked out in detail.
Another area for further work is the dynamics of chiral
symmetry breaking.  Particularly worthy of attention is the
possibility that the Higgs fields themselves have a
dynamical origin, i.e., that the spontaneous $G$-symmetry
breaking and the chiral symmetry breaking both arise from
some sort of Nambu-Jona-Lasinio mechanism involving the
$L$-fermions.  It is nevertheless encouraging that the
purely group-theoretic and kinematic foundation of such an
enterprise, as described in this paper, has proved to have
no serious drawbacks.

\section{Universality Properties of Fermions}

\noindent
{\bf 2A. Implementing Universality}

Considering all leptons and quarks together, a fermion is
labelled by a family index $i=1,\ldots ,N$, $N$ assumed
arbitrary, a flavour index $f = 1,2$ ( 1 is ``up'' and 2 is
``down'', say) and a generalised colour index $\alpha =
0,1,2,3$, $\alpha=0$ referring to leptons and $\alpha = 1,2,
3$ (collectively denoted by $c$ where necessary) to
conventional quark colour.  The $8N$ dimensional complex
vector space $V$ spanned by the orthonormal basis $\{ |i,
f,\alpha \rangle \}$ is the space of 1-fermion states
(ignoring momenta and helicities).  In any gauge model for
the interactions of all the 
fermions, $V$ will carry a unitary, not necessarily
irreducible, representation of the global gauge group.  The
chirality assigned to the vectors of $V$ (e.g., whether
Dirac or Weyl spinors) will determine the parity properties
of the various currents and is left open for the moment. It
will turn out that the only viable choice is in favour of
left-handed Weyl spinors; in fact it is a unique and
nontrivial feature of the model that right-handed fermions
cannot exist in the gauge - invariant  limit of the model;
they will be generated spontaneously (along with masses) in a
very natural way.

The basic strategy pursued in this paper is to postulate
first that underlying the standard $G = SU(2) \times U(1)
\times SU(3))$ gauge model there is a unifying or
embedding gauge theory with group $\cg$, treating all
fermions on an equal footing and, then, to determine $\cg$
as the group having the property that its subgroup
respecting all observed universality properties is $G$.  In
the actual execution, it is simpler to proceed by choosing $\cg$ as the
smallest simple group that will suffice, namely $\cg = SU
 (8N)$, and then to verify that its subgroup satisfying
weak and strong universality is indeed $G$.  The universality
properties referred to are those valid at the level of $SU(2)
\times U(1) \times SU(3)$, before it is spontaneously broken
to the $SU(3) \times U(1)$ model incorporating fermion mass
differences and family mixings.

We denote a basis for the Lie algebra of $\cg = SU (8N)$
by $\{ t_A,A=1,\ldots,64N^2-1\}$ and write the fermion
current coupling to the $\cg$-gauge boson $X_A$ as 
$$J_A = \ol{\psi}t_A \psi$$
where $\psi \in V$ and the space-time structure has been
suppressed.  Let us choose the index $A$ to be compatible with
the family, flavour and colour labels as they occur in the
$G$ gauge theory: $A$ is then a pair of sets of indices
$(i,f,\alpha )$ and $(i',f',\alpha')$.  A typical
universality property is most simply formulated as the
statement that the current $J$ which couples to a
particular gauge boson $X$ of the $G$ gauge theory is
unchanged by a particular set of permutations of the labels
$(i,f,\alpha )$ of a fermion.  Weak universality of leptonic
currents, for instance, is the generalisation of the old
$e-\mu$ universality to the statement that the charged weak
currents are invariant under a permutation of $e$, $\mu$ and
$\tau$ and the {\it same} permutation, simultaneously, of
$\nu_e, \nu_{\mu}$ and $\nu_{\tau}$.  Postponing a more precise
formulation of universality to the next two subsections, we
note first that a universality transformation is a unitary
matrix on $V$ and that the set of universality
transformations leaving invariant a given subset of the
currents is a group.

Given a group $S$ of unitary operators on $V$, a $\cg$-gauge
fermion current $J = \ol{\psi} t \psi$, where $t$ is a real
linear combination of $\{ t_{A}\}$, is invariant under $S$
(``universal with respect to $S$'') if 
$$\ol{\psi} s^* t s \psi \equiv \ol{\psi}s^{-1}t s\psi =
\ol{\psi} t \psi$$
for all $s \in S$.  The set of $t$ which commute with every
$s \in S$
forms a Lie subalgebra of the Lie algebra of $\cg$ and the
corresponding Lie group is a subgroup of $\cg$, the {\it
centraliser} of $S$ in $\cg$, denoted $C(S)$.  The gauge 
bosons of the $C(S)$ gauge theory couple to all currents
universal with respect to $S$ and only to them, and we
obtain in this way an embedding of the universal gauge
theory in the $\cg$ gauge theory.  The ratios of the gauge
couplings of each simple factor group of $C(S)$, as well as
the couplings of various fermion currents to any abelian
factor group of $C(S)$, are thereby fixed.  Thus one of the
prime motivations for unification is fulfilled by appealing
to the observed pattern of families rather than by invoking
complicated Higgs multiplets.  Indeed, though the Higgs
mechanism for reducing $\cg$ to $G$ certainly remains a
viable option, we are now free to explore other, physically
more compelling, reasons for the manifestation of
universality and family structure.  One such speculative
possibility will be suggested at the end.

\noindent
{\bf 2B. Weak Universality}

As already stated above, we follow conventional wisdom and
begin by supposing that the weak interactions of leptons in
the unbroken standard $G$ gauge theory are unchanged by an
arbitrary permutation $\pi$ of the family index of the up
leptons and the same permutation of the down family index.
$$|i,u,\alpha=0 \rangle \ra |\pi (i), u,\alpha=0\rangle , ~|i,
d,\alpha =0\rangle \ra |\pi (i), d, \alpha =0 \rangle
.$$
The physical leptons may violate strict universality
through mass differences and possible family mixings; both of
these are generally taken to be manifestations of the
breaking of $G$ further to the final, low energy, exact
gauge group $U(1)_{em} \times SU(3)$.  Under the same
assumption, the weak interactions of quarks in the $G$-gauge
theory are also invariant under similar simultaneous
permutations of the up and down family indices (despite the
family mixings present in low energy currents).  Hence the
{\it conventional} universality assumption can be stated as
the invariance of all weak interactions under 
$$|i,u,\alpha \rangle \ra |\pi (i),u,\alpha \rangle
,~~|i,d,\alpha \rangle \ra |\pi (i) ,d, \alpha \rangle \eqno
(1)$$
simultaneously for each fixed $\alpha = 0,1,2,3$.

However, this formulation of weak universality is
incomplete.  Firstly, if we ignore, once again, family
mixings (i.e., at the $G$-invariant stage), the observed
weak currents of a definite colour are of the general forms
$\sum_i \ol{u}_{i\alpha} \Gamma d_{i\alpha}$ (and its
conjugate), $\sum \ol{u}_{i\alpha} \Gamma u_{i\alpha}$ 
 and $\sum
\ol{d}_{i\alpha} \Gamma d_{i\alpha}$ where $\Gamma$ are
(different) matrices which {\it do not} operate on $i$.  So
a unitary transformation of the family index $i$ will
commute with $\Gamma$; the currents are invariant not just
under the discrete permutations $\pi$ but the more general
$$|i ,u,\alpha \rangle \ra U_{ij} | j,u,\alpha \rangle
,~~|i,d,\alpha \rangle \ra U_{ij} |j ,d,\alpha \rangle
\eqno (2)$$
where $U$ in both transformations is the same unitary matrix
[11].  Next, as far as the weak currents are concerned,
quark colour is no different from the family label; the
coupling strengths of the currents do not depend on the
quark colour and the matrices $\Gamma$ do not operate on
them.  We may therefore immediately extend weak universality
to encompass unitary transformations on the pair of indices
($i$, $c=1,2,3$).  Moreover the only empirical reason for
not extending it to include leptons also would appear to be
the lack of equality of the coupling strengths of leptons
and quarks to the photon and the $Z$ boson.  The couplings
of the physical $\gamma $ and $Z$ to fermions are however
fixed only after $SU(2)\times U(1) \subset G$ invariance is
broken down to $U(1)_{em}$ invariance and the neutral bosons
mixed and hence this apparent reason is not so compelling.
This leaves the theoretical objection that the so called
weak hypercharges in the standard $SU(2) \times U(1)$ model
does distinguish between quarks and leptons.  The remarkable
fact is, as will become clear in the next subsection, that
in our unification scheme the hypercharge is not an
attribute of weak but of strong universality and is a
measure of the absence of strong interactions for the 
leptons.

In the light of the above considerations, we take the
maximal weak universality group $S_{wk}$ as the group 
consisting of unitary transformations with matrix elements
$U_{i\alpha ,j\beta}$, $i,j=1,\ldots ,N$, $\alpha ,\beta
= 0,1,2,3$, applied to the fermion basis $|i,f,\alpha
\rangle$ leaving the flavour $f$ unchanged:
$$\begin{array}{l}
|i,u,\alpha \rangle \ra U_{i\alpha ,j\beta} |j,u,\beta
\rangle , \\
|i,d,\alpha \rangle \ra U_{i\alpha ,j\beta} |j,d,\beta
\rangle .
\end{array} \eqno (3)$$
Abstractly, $S_{wk}$ is the group $U(4N)$.  If we write $V$
as a tensor product of spaces spanned by family, flavour and
colour basis vectors, $V = V_{fam} \otimes V_{fl} \otimes
V_{col}$, $S_{wk}$ is the unitary group of $V_{fam} \otimes
V_{col}$.  In this abstract sense, weak universality is just
the statement that, as far as weak interactions are
concerned, family and colour directions in $V$ can be chosen
as an arbitrary orthonormal basis for $V_{fam} \otimes
V_{col}$.  

In accordance with the general considerations of section
$2A$, the subgroup of $\cg = SU(8N)$ whose currents respect
weak universality is the centraliser $C(S_{wk})$.  As a $(8N
\times 8N)$ matrix group on $V$, $S_{wk}$ consists of pairs
of identical $U(4N)$ matrices $s_{wk}$ acting on $V_{fam}
\otimes V_{col}$:
$$s_{wk} = \pmatrix{x_{wk}& 0 \cr 0 & x_{wk}}; \eqno (4)$$
i.e., $S_{wk}$ is the diagonal subgroup $\{ (x_{wk}, x_{wk})
\}$ of the direct product $U(4N) \times U(4N)$.  We write
this as
$$S_{wk} = D^2U(4N), \eqno (5)$$
so that 
$$G_{wk} = C(D^2U(4N)). \eqno (6)$$
It is very easy to determine the group $G_{wk}$.  In the
basis used in Eq. (4) (i.e., a fixed flavour labelling the
first $4N$ rows and columns), write a general element $g \in
SU(8N)$ as the matrix
$$g = \pmatrix{g_{uu} & g_{ud} \cr g_{du} & g_{dd}}, $$
with each $g_{ff'}$ a $4N \times 4N$ matrix.  For $g$ to
commute with $s_{wk}$, each submatrix $g_{ff'}$ must commute
with $x_{wk}$ and since $x_{wk}$ is an arbitrary matrix,
each $g_{ff'}$ is trivial:
$$g_{ff'} = a_{ff'} 1_{4N},$$
where $a_{ff'}$ are complex numbers (the subscript on 1
indicating dimension will prove its usefulness soon).
Hence a general element of $G_{wk}$ is of the form 
$$g_{wk} = \pmatrix{a_{uu}1_{4N} & a_{ud}1_{4N} \cr
a_{du}1_{4N} & a_{dd}1_{4N}}, \eqno (7)$$
$a_{ff'}$ being arbitrary complex numbers such that $g_{wk}$
is unitary and has determinant 1.  A reordering of the basis
vectors of $V$ (interchange the $(n+1)$th row (column) with
the $(4N+n)$th row (column) of the matrix) allows us to write
a typical element of $G_{wk}$ as the $8N \times 8N$ matrix
having the unitary matrix
$$a = \pmatrix{a_{uu} & a_{ud} \cr a_{du} & a_{dd}}$$
repeated $4N$ times along the principal diagonal.  Thus
$G_{wk}$ is the unit determinant diagonal subgroup of
$U(2)^{4N}$: 
$$G_{wk} = SD^{4N} U(2). \eqno (8)$$

The condition $\det g_{wk} = (\det a )^{4N} = 1$ allows us
to rewrite this as the diagonal group $D^{4N}U(2)_{4N}$,
isomorphic to $U(2)_{4N}$, the group of unitary matrices
having determinant equal to any $4N$th root of unity.  This
is obviously not a connected Lie group; its connected
component is $SU(2)$.

Since the currents in a gauge theory are completely
specified in terms of the Lie algebra of the gauge group,
possible lack of connectedness (and simple-connectedness)
of the group is immaterial in the (perturbative) calculation
of any process.  Nonperturbative results may depend on the
topology of the group, but such considerations are not
pursued in this paper.  With this qualification, we have
thus shown that
$$G_{wk} = SU(2). \eqno (9)$$

It is immediately evident that the $8N$ basis vectors
$\{|i,f,\alpha \rangle \}$ of the fundamental fermion
representation of $SU(8N)$ form $4N$ copies of the
fundamental representation of $SU(2)$, the $SU(2)$ acting on
flavour and the copies labelled by family and colour.  The
relationship between weak universality and family structure
has thus been made precise and concrete by the use of
$SU(8N)$ as an embedding or unifying group.

\noindent
{\bf 2C. Strong Universality}

The formulation of a strong universality principle for
quarks proceeds in a manner very similar to that of weak
universality: the colour currents of quarks responsible for
strong interactions are universal in the sense that they are
sums over family and flavour, with a common coupling
constant.  This feature is usually recognised as being the
reason for the phenomenon loosely called flavour invariance
(approximate because of quark masses).  Exactly as above,
the strong universality group is then $D^3 U (2N)$.
Ignoring the existence of leptons for a moment and taking
the primitive gauge group to be $SU(6N)$, the resulting
strong gauge group would then be $S(D^{2N}U(3))$.

But leptons do exist and the fact that they have no strong
interaction -- i.e., no (low energy) current changes leptonic
colour to any other colour -- introduces a fundamental new
feature.  In the context of universality this means that the
strong interactions are impervious to any arbitrary choice
of basis of the $2N$ dimensional leptonic subspace of $V$
spanned by $\{ |i,f,\alpha =0\rangle \}$, completely
independent of the universality transformations of the quark
subspace.  The most general strong universality
transformation, as a matrix on $V$, is therefore of the form
$$s_{st} = \pmatrix{x_{st}' & & & \cr & x_{st} & & \cr & &
x_{st} & \cr & & & x_{st}}, \eqno (10)$$
with $x_{st}' \in U(2N)$ and $x_{st} \in U(2N)$, all blank
entries being zero.  (The basis used here is, of course, one
in which the first $2N$ rows and columns correspond to
leptons of all family and flavour).  The strong universality
group is thus
$$S_{st} = U(2N)\times D^3U(2N). \eqno (11)$$

As before, the strong gauge group $G_{st}$ is the centraliser
of $S_{st}$ in $SU(8N)$.  To compute it, write $g \in
SU(8N)$ as the matrix (in the basis used in Eg. (10)):
$$g = \pmatrix{g_{00} & \cdots & g_{03} \cr \vdots & & \cr g_{30} &
\cdots & g_{33}} \equiv (g_{\alpha\alpha'})$$
each $g_{\alpha\alpha'}$ being a matrix on $V_{fam} \otimes
V_{fl}$.  For $g$ to commute with $s_{st}$, we must have
$$\begin{array}{l}
g_{00}x_{st}' = x_{st}'g_{00}, ~~ g_{cc}x_{st} =
x_{st}g_{cc}, \\
g_{c0}x_{st}' = x_{st}g_{c0}, ~~ g_{oc} x_{st} =
x_{st}'g_{oc}, \\
g_{cc'}x_{st} = x_{st} g_{cc'}~~\mbox{for}~ c \neq c'\\
\end{array}$$
for all $x_{st}$, $x_{st}' \in U(2N)$ and $c,c' = 1,2,3$.
These conditions are solved by
$$g_{00} = b_{00} 1_{2N}, ~~g_{c0} = g_{0c} = 0,~~ g_{cc'} =
b_{cc'} 1_{2N}~ \mbox{for all}~ c,c',$$
for arbitrary complex numbers $b_{00}$ and $\{b_{cc'}\}$.
Reordering rows and columns, a typical element of $G_{st}$
can therefore be written as the $8N \times 8N$ unitary
matrix having the $4 \times 4$ matrix
$$b = \pmatrix{b_{(l)} & 0 \cr 0 & b_{(q)}},$$
where $b_{(l)}$ is an arbitrary complex number with
$|b_{(l)}|=1$ and $b_{(q)}$ is a unitary $3 \times 3$
matrix, repeated $2N$ times along the diagonal.  The matrix
$b$ operates on $V_{col}$ whose leptonic and quark subspaces
are distinguished by the subscripts.

Thus, as a $(8N\times 8N)$ matrix group, the strong gauge
group is 
$$G_{st} = SD^{2N} (U(1) \times U(3)). \eqno (12)$$
The unit determinant condition says that 
$b_{(l)}^{2N} (\det b_{(q)})^{2N} =1$; this puts no
restriction on $\det b_{(q)}$ but merely says that its value
is fixed in terms of the arbitrary phase $b_{(l)}$.  Hence
$G_{st}$ is connected and is isomorphic to $U(3)$.
Restricting ourselves to the Lie algebra and ignoring
topological niceties once again, we therefore write
$$G_{st} = U(1) \times SU(3). \eqno (13)$$
It is clear that the $SU(3)$ group operates 
trivially on the subspace $\alpha =0$ (leptons) and as the
fundamental {\bf 3} representation on the quark subspace of
each family and flavour; the total fermion space $V$ breaks
up as $2N$ copies of the colour triplet and $2N$ copies of
the singlet representation.  The $U(1)$ subgroup is not just
a group acting on leptons, as it arises from solving the
unit determinant condition.  Its physical interpretation
will become clear in the next section.

\section{The Standard Model}

{\bf 3A. Combining $G_{wk}$ and $G_{st}$}

Physically, the picture we would like to have is this:  the
effective gauge theory arrived at by appealing to
universality should accommodate all currents satisfying {\it
either} weak universality {\it or} strong universality and
no others.  This will be ensured if we can conclude that the
effective gauge group $G$ is just the direct product of 
$G_{wk}$ and $G_{st}$ and that, in turn, requires that
$G_{wk}$ and $G_{st}$ are disjoint as subgroups of $\cg$ and
that they are mutually commutative.  These two properties are
immediately obvious for the Lie algebras of $G_{wk}$ and
$G_{st}$ (e.g., colour generators and flavour generators
commute) and hence for the groups themselves if they are
connected and simply connected.  Since we have already
chosen to confine attention to the Lie algebra we may assert
that the subgroup of $\cg = SU (8N)$ describing the
effective gauge theory which respects {\it either} weak {\it
or} strong universality is indeed
$$G = G_{wk} \times G_{st} = SU(2) \times U(1) \times
SU(3). \eqno (14)$$
Under the $SU(2) \times SU(3)$ subgroup of $G$, leptons of
each family transform as the representation $(\botwo ,\bone)$ and
quarks of each family as the representation $(\botwo ,\bothree)$.
($m$ and $n$ in the notation $(\bfm ,\bfn)$ are the
dimensions of $SU(2)$ and $SU(3)$ representations).  The
$U(1)$ transformation properties of (i.e., the $U(1)$
coupling constants to) the various fermions will be
determined in the next subsection -- it will turn out that
$U(1)$ is in fact the group of what is conventionally called
the weak hypercharge.

At one level one may think of the results of this section as
just a systematisation of the relationship between observed
universality and family structure.  Some striking insights
have nevertheless emerged in the process:

\begin{enumerate}
\item If, as we have hypothesised, the family structure (the
$N$-fold replication of an irreducible representation) of
fermions in the standard model is the result of imposing
universality restrictions on the currents of a unifying
gauge group $\cg$, the standard gauge group $G = SU(2)
\times U(1) \times SU(3)$ fixes $\cg$ {\it uniquely} to
be $SU (8N)$.  The number of families plays no role in the
reduction of $\cg$ to $G$ (except in so far as the
topological properties of $G$ are concerned).

\item The hypercharge $U(1)$ actually arises from strong
universality and reflects the absence of strong interactions
among the leptons.

\item The fact that weak $SU(2)$ currents are left-chiral
forces all fermions to be left-handed  Weyl spinors in the
primitive $\cg$ gauge theory.  Obtaining  a $G$ gauge theory in
which the $SU(2)$ currents are $V-A$ while the $SU(3)$ currents
are pure $V$ then poses a problem.  
\end{enumerate}

The usual ways of
introducing $R$-fermions in a ``grand'' unified model
broadly fall into two distinct strategies: 
i) They belong to a different representation of a
simple unifying group $\cg$ chosen carefully so that when
$\cg$ is broken to $G$, the parity properties of currents
come out right.  The prototype of this method is of course
the $SU(5)$ model [12] in which this is done for each family
separately.
ii) They transform trivially under $\cg$ but
nontrivially under another subgroup $\cg'$ 
of a full unifying (nonsimple) group $\cg \times \cg' \equiv
\cg_L \times \cg_R$.  The gauge bosons of the standard model
then result from a mixing of the bosons  of $\cg$ and
$\cg'$, induced by spontaneous symmetry breaking.  The
prototypes here are left-right symmetric models [13].  Such
strategies are not natural for us.  For the first option to
work, the representation of $SU(8N)$ to which $R$-fermions
are to be assigned must evidently have dimension less
than $8N$ and there are no such.  The second option is
excluded by universality itself.  Observed universality does
not discriminate between $V-A$ (charged weak) and $V$
(electromagnetic and colour) currents: if
we apply universality to, say, $\cg_L \times \cg_R$ as the
unifying group, we shall end up with $G_L \times G_R$ as the
effective group, requiring further (Higgs?) gymnastics to
end up with the correct $V-A$ and $V$ currents.

Thus, accepting universality as the sole guide in the choice
of the unifying group forces us to look for novel ways of
understanding the parity properties of currents.  A possible
solution of this problem is described in section 4B.

\noindent
{\bf 3B. Coupling Constants}

The ratios of the squares of the coupling constants of
$SU(2)$ and $SU(3)$ currents and of each individual fermion
$U(1)$ current are determined by the way $G$ is embedded in
$\cg$.  (It is useful to remember that the physical
mechanism of symmetry reduction is immaterial for this
purpose).  Fix a normalisation 
$${\rm tr}~ t_A^2 = \delta \eqno (15)$$
for the generators of $SU(8N)$.  The generators of the
$SU(2)$ subgroup can then be written, as $SU(8N)$ generators,
as the matrices
$$T_a = \frac{1}{2} g ~{\rm diag}~ (\tau_a, \tau_a, \ldots
,\tau_a), ~~a =1,2,3,$$
in the flavour basis, where $\tau_a$ are the Pauli matrices
(${\rm tr}~ \tau_a^2=2)$, i.e., as the matrix having $\tau_a$
repeated $4N$ times along the diagonal and all other entries
zero.  The normalisation (15) fixes the value of $g^2$:
$$\delta = {\rm tr}~ T_a^2 = \frac{1}{4} g^2 4N {\rm tr}~ \tau_a^2 =
2Ng^2,$$
i.e.,
$$g^2 = \frac{\delta}{2N}. \eqno (16)$$
Similarly we write the generators of the $SU(3)$ subgroup as

$$L_n = \frac{1}{2} f ~{\rm diag}~ (0,\lambda_n,0,\lambda_n,\ldots
,0,\lambda_n), ~~ n=1,\ldots ,8,$$
in the colour basis, where $\lambda_n$ are the Gell-Mann
matrices $({\rm tr}~ \lambda_n^2 =2)$, with $(0,\lambda_n)$
repeated $2N$ times along the diagonal.  So
$$f^2 = \frac{\delta}{N} = 2g^2. \eqno (17)$$
The $U(1)$ group commutes with $SU(2)$ and $SU(3)$ and hence
the hypercharge $Y'$ as a generator of $SU(8N)$ commutes
with $\{ T_a\}$ and $\{ L_n\}$.  Therefore
$$Y' = \frac{1}{2} ~{\rm diag}~ (g'1_{2N},g'' 1_{6N})$$
(for real numbers $g'$ and $g''$) in the flavour basis.
Equivalently, in the colour basis,
$$Y' = \frac{1}{2} ~{\rm diag}~ (g',g''1_3,\ldots ,g',g''1_3)$$
with $2N$ repetitions.  Since $Y'$ is traceless,
$$g'' = - \frac{1}{3} g' \eqno (18)$$
and we write, to conform to usual practice, 
$$Y' = \frac{1}{2} g' Y, ~~Y = {\rm diag}~\left(-1_{2N},\frac{1}{3}
1_{6N}\right )$$
in the flavour basis.  The correct ratio of the lepton and
quark hypercharges follows strictly from the simplicity of
$SU(8N)$ and the fact that there are three colours of quarks
and one of leptons (it is not directly related to baryon
and lepton numbers).  The common coupling constant $g'$ is
fixed by the normalisation $\delta = {\rm tr}~Y'^2$ to be
$$g'^2 = \frac{3\delta}{2N} = 3g^2. \eqno (19)$$
The ratios of the coupling constants are different from the
almost canonical values expected [14] in a sequential
embedding of the standard model in any simple unifying
group with fermions in the $(\botwo,\bone) \oplus (\botwo,\bothree)$
representation of $SU(2) \times SU(3)$ because there are no
$R$-fermions in our model.  (Indeed, since these ratios are
fixed by the way $G$ sits inside $SU(8N)$, this is yet
another indication of the impossibility of incorporating
$R$-fermions in any representation of $SU(8N)$ without
destroying the standard model).  Consequently, the values of
the mixing angle and the strong coupling constant are both
twice their canonical values:
$$\sin^2\theta = \frac{3}{4}, \hspace{1cm}~~
\frac{\alpha_S}{\alpha_W} = \frac{f^2}{g^2} = 2. \eqno
(20)$$
The first of these numbers, in particular, shows that the
unification regime of the model has to be at substantially
higher energy then we have been used to (see later).

\section{Breaking $G$ Invariance}

{\bf 4A. The Choice of Higgs Fields}

So far we have avoided specifying the mechanism that is
responsible for reducing the gauge symmetry from the
primitive (``would-be'') gauge group $\cg$ to the effective
one of the standard model.  In our model this mechanism is
no different from that responsible for the universality
properties of the $G$-gauge theory; hence, it is possible to
think that looking for the underlying physical causes of
universality might offer options other than the search for an
astute Higgs representation of $\cg$.  We postpone this
question to the concluding section.

In contrast, the breaking of $G$ to $SU(3) \times U(1)_{em}$
is attributed in this paper to a conventional Higgs
mechanism.  The main reason is that our real concern is how to
understand the (pre-Higgs) standard model and its observed
systematics in terms of a unified model, rather than in the
details of how it is further broken.   In any case, the
standard Higgs mechanism is extraordinarily successful
compared to the many efforts to supplant it or to find a more
fundamental explanation for it.  At the same time, it is
very simple -- compare the complicated Higgs multiplets
required in most (one-family) grand unified models.  An
additional reason is the central role the Higgs fields play
in our approach to chiral symmetry breaking and in
generating $R$-fermions.  (See, however, the remark at the
end of the Introduction).

Accordingly, we assume that the $\cg$ gauge theory has one
Higgs field $\Phi$ transforming as the defining
(fundamental) representation of $SU(8N)$.  With respect to
$G \subset \cg$, $\Phi$ decomposes exactly as the fermion
representation: there are $N$ families $\phi_i$, each
$\phi_i$ transforming under $SU(2) \times SU(3)$ as
$(\botwo,\bone) \oplus (\botwo,\bothree)$ with $Y = -1$ and $\frac{1}{3}$
respectively (lepton-like and quark-like Higgs).  We assume
further that only the up flavours of lepton-like Higgs of all
families condense in the vacuum, with vacuum expectation
values $\eta_i$.  Except for family replication, this is the
standard Higgs picture as far as $SU(2) \times U(1)$ is
concerned.  But the tree level algebraic properties of the
standard model -- the relations involving various observables
like the values of the photon and $Z$ coupling constants and
the mixing angle $\theta$, the masses of $W$ and $Z$, etc. --
are insensitive to the number of Higgs doublets (having
vacuum expectation values all in the same flavour).  Thus
the fully broken $SU(2) \times U(1)$ theory is
indistinguishable from the standard model except in higher
orders involving Higgs propagators (and also of course when external
Higgs particles are involved).  We note in particular that
the electric charges of all the fermions (so far, only
left-handed) have the conventional values.

In addition, we have the colour triplets of quark-like
Higgs (which may or may not be massive).  They are presumably
confined and cannot have $SU(3)$ invariant couplings to
quarks -- indeed, $\Phi$ has no $SU(8N)$ invariant coupling
to the fermion $\boeign$.  They can interact with gluons
however and so will form part of the ``sea'' in hadrons.

\noindent
{\bf 4B. Right-handed Fermions}

We have already noted (Sec. 3A) the difficulty of
introducing $R$-fermions in the $\cg$ gauge model in such a
way as to insure, in the effective $G$ gauge theory, that
$SU(2)$ currents are $V-A$ while the $SU(3)$ currents are
$V$.  As a way out of this difficulty, the proposal put
forward here is that, while $R$-fermions are absent in the
$\cg$ gauge model, they are generated dynamically,
simultaneously with the spontaneous breaking of $G$ to
$SU(3) \times U(1)_{em}$.  In support of the general idea
that $R$-fermions might really be the result of the final
stage of symmetry breaking (of $G$), we point to two very
suggestive circumstances: 
i) They are essential in restoring the parity
invariance of strong and electromagnetic interactions,
precisely those interactions whose gauge invariance
survives all symmetry breaking; in other words, those and
only those currents which couple to massless gauge bosons
are pure $V$. 
ii) They are necessary to generate fermion masses
which are also a manifestation of symmetry breaking.  The
only phenomena which require $R$-fermions are thus closely
linked to $G$-symmetry breaking.

Explicitly, our proposal is that $R$-fermions are composites
of $L$-fermions and Higgs bosons in certain specific
channels.  First, the binding must be orbitally excited (to
obtain the chirality flip) and must be, in a first
approximation [15], within the same family (to preserve the
universality properties of realistic Dirac fermions).  Dropping
the family index and writing $\psi_L$ for left handed Weyl
spinor fields of a given $SU(2) \times SU(3)$ multiplet, the
composite field $\psi_L\partial_{\mu} \phi$ transforms under
the Lorentz group as $(\frac{1}{2},0) \otimes
(\frac{1}{2},\frac{1}{2}) = (1,\frac{1}{2}) \oplus
(0,\frac{1}{2})$ of which the right-handed
$(0,\frac{1}{2})$ component can be projected out by taking 
$$\psi_R \simeq \sigma_{\mu} \psi_L \partial_{\mu} \phi
\eqno (21)$$
where $\sigma_0 =1$ and $\sigma_i$ are the usual spin
matrices.  Evidently this equation can have, at this stage,
only a schematic  meaning, as a means of keeping track of
quantum members.  At the very least, we have ignored the
need for proper regularisations of quantum composite
fields.  The dynamical problems which will have to be solved
in implementing this idea cannot obviously be broached
here.  Nevertheless, Eq. (21) can already be seen to have an
intriguing consequence.  On writing
$$\psi_R \simeq \partial_{\mu} (\sigma_{\mu}\psi_L \phi ) -
(\partial_{\mu}\sigma_{\mu}\psi_L)\phi,$$
it is clear that if both $\psi_L$ and $\psi_L\phi$ are
($L$-projections of) massless Dirac fields, $\psi_R$ vanishes;
right handed components can be generated dynamically in this
way only if
mass is also simultaneously generated.  In the rest of this
section, we shall disregard all dynamical problems
concentrating only on the quantum number aspects.

In general $\psi_L\phi$ composites can transform under
$SU(2) \times SU(3)$ as any of the irreducible
representations occurring in the decomposition of $(\botwo ,\bothree
\oplus \bone ) \otimes (\botwo ,\bothree \oplus \bone)$.  To delimit the
number of possible irreducible fields, let us assume
that $\psi_L\phi$ binding takes place for exactly the same
irreducible representation (with respect to $G$) of $\phi$
which also condenses in the vacuum, namely the lepton-like
(colourless) representation $\phi \sim (\botwo ,\bone)$, and, furthermore, in
the antisymmetric tensor product $\wedge$ of $SU(2)$
representations.  Then
$$\begin{array}{lcl}
\psi_L \phi & \sim & ( \botwo \wedge \botwo , (\bothree \oplus \bone )
\otimes \bone ) \\
& = & (\bone ,\bothree) \oplus (\bone ,\bone ).
\end{array}$$

We now identify the two irreducible representations on the
right as the down components of $R$-quarks and $R$-leptons respectively.
Explicitly, in a general gauge, 
$$\psi_{Rd}(\bothree ) \simeq \psi_L (u,\bothree ) \phi (d,\bone ) -
\psi_L (d,\bothree) \phi (u,\bone) \eqno(22)$$
and
$$\psi_{Rd}(\bone ) \simeq \psi_L (u,\bone ) \phi (d,\bone ) -
\psi_L (d,\bone) \phi (u,\bone) \eqno (23)$$
The up $R$-fermions are correspondingly obtained by
replacing $\phi$ by its conjugate, $\phi^c$:
$$\psi_{Ru} (\bothree) \simeq \psi_L (u,\bothree ) \phi^c (d,\bone ) -
\psi_L(d,\bothree)\phi^c (u,\bone ), \eqno (24)$$
$$\psi_{Ru} (\bone) \simeq \psi_L (u,\bone ) \phi^c (d,\bone ) -
\psi_L (d,\bone ) \phi^c (u,\bone ). \eqno (25)$$
The subscripts $u$ and $d$ on $R$-fermions (as distinct from
$L$-fermions for which they are written as arguments) are
meant only to indicate in advance which composite will turn
out to be the $R$-partner of which $L$-fermion in the
currents coupling to $U(1) \times SU(3)$ -- they are all
invariant under $SU(2)$.   The totality of $R$-fermions falls
into two sets of quarks $\psi_{Ru} (\bothree)$ and $\psi_{Rd}
(\bothree)$ and two sets of leptons $\psi_{Ru} (\bone )$ and
$\psi_{Rd} (\bone)$ in each family.  This is exactly the
pattern required by the standard model.

Finally, to fix the parity properties of the total fermionic
currents of $U(1)$ and $SU(3)$, we need to know the
corresponding coupling constants of the $R$-fermions.  The
transformation properties of $\psi_R$ under $SU(2) \times
SU(3)$ immediately imply that no $R$-currents couple to
$SU(2)$ and no leptonic $R$ currents couple to $SU(3)$.  The
$R$-quark coupling to $SU(3)$ is determined, just by $SU(3)$
gauge invariance, to be the same as the $L$-quark coupling
($=$ the intrinsic gauge coupling constant of $SU(3)$).
Hence all $SU(3)$ currents are pure $V$.  The fermion
couplings to abelian groups such as the hypercharge $U(1)$
or (equivalently) the electromagnetic $U(1)$ are not fixed
in this way as there is no intrinsic gauge coupling defined
by the gauge field; they have to be determined for each
fermion individually.  But being abelian charges, they are
additive in each composite and we may compute them directly
and easily.  For the electric charge, we get
$$Q(\psi_{Rd} (\bothree )) = \left \{
\begin{array}{l}
Q(\psi_L(u,\bothree))+Q(\phi (d,\bone )) = \ds \frac{2}{3} -1 \\
Q (\psi_L (d,\bothree )) +Q (\phi (u,\bone )) = \ds - \frac{1}{3} +
0 \end{array} \right \} = - \frac{1}{3}$$
and, similarly,
$$\begin{array}{rcl}
Q(\psi_{Rd} (\bone )) & = & -1, \\
Q(\psi_{Ru} (\bothree )) &= &\ds \frac{2}{3}\\
Q(\psi_{Ru} (\bone)) &=& 0. \\
\end{array}$$
Hence the electromagnetic current is also pure $V$.  It is
pertinent to stress that the electric charges of $\psi_L$
and $\phi$ are themselves fixed completely by the embedding
of the hypercharge in $SU(8N)$ and by the Higgs structure of
the model; they are not assigned {\it a priori}.

Thus the assumption that there exists an attractive force between
$L$-fermions and Higgs in a specific ``channel''
corresponding to unique angular momentum, colour and flavour
selection rules, strong enough to bind, leads to exactly the
right global quantum numbers for the $R$-fermions.  To make
the picture complete, this is of course not enough.  One
needs to be able to construct local field operators for
$\psi_R$ such that they transform correctly also under {\it
local} gauge transformations, in other words, deal with the
dynamical problems alluded to earlier.  In conventional
(unconstrained) quantum field theories, the way to
construct local field operators for composites has been
known since 
long [16], but its generalisation to gauge theories remains
an open problem.

\section{Open Questions}

Of the questions we have so far left unaddressed, the most
pressing is that of the physical mechanism which reduces the
primitive gauge group $SU(8N)$ to $SU(2) \times U(1) \times
SU(3)$ and, conversely, the significance and regime of
validity of full $SU(8N)$ gauge invariance.  It may be
reassuring to note at the outset that it is possible to find
a Higgs representation which will serve the propose.  It is
shown in the Appendix that when a gauge group $\cg$ is
spontaneously broken to a subgroup $G$ which is the
centraliser of a group $S \subset \cg$, we can always find a
set of Higgs fields belonging to the adjoint representation
of $S$ and a set of non-zero vacuum values for them such
that their little group (stabiliser) is precisely $G$; it is
also shown there that two copies of the adjoint
representation of the family $SU(N)$ group, suitably
embedded in the adjoint representation of $SU(8N)$, are
sufficient to break $SU(8N)$ down to $SU(2) \times U(1)
\times SU(3)$.

If the Higgs option is chosen, then conventional wisdom
dictates that the vacuum  values of the adjoint Higgs will
determine the energy scale at which unification will hold
and all $64N^2-1$ currents of $SU(8N)$ will be operative.
The relatively large values of $\sin^2\theta$ and of
$f^2/g^2$ that we have found in section 3B indicate that
this unification energy (and hence the masses of the exotic
(non-universal) gauge bosons of $SU(8N)$) will be many orders
of magnitude larger than the $10^{15} GeV$ or so that we
are so used to and, even, the Planck mass.  Conceptually,
this is unknown teritory; in any case no reasonable physical
sense can be attached to a procedure of evolving
low energy parameters beyond the Planck mass while ignoring 
gravitational effects in the renormalisation group equations.

The other question touching on the magnitude of the
unification energy scale is that of anomalies.  Though the
$SU(8N)$ gauge theory is not anomaly-free,  our low energy
``universal'' model, being indistinguishable from the
standard model, is:  the dynamically generated $R$-fermions
cancel precisely the triangle anomalies arising from the
$L$-fermions as long as no non-standard gauge boson internal
lines occur in loops.  Such loops will technically be
non-renormalisable; however,  their contribution will be negligibly
small in any regularised computation, as long as the cut-off
is much smaller than the unification scale.  This is just a
loose paraphrase of the set of results collectively known as
decoupling theorems.  Since gravitational effects
effectively preclude a cut-off above the Planck mass, we need
not take the demand of technical renormalisability too
seriously at this stage.  In any case, these points are only
of academic interest if, as we suggest below, the reduction
$\cg \lr G$ is caused by a 
mechanism other than that of Higgs fields.

The option that we favour is to take leptons and quarks to
be composites of a set of preons (or metaparticles) and to
attribute the distinct gauge interactions of each fermion to
the preons which themselves have the same gauge
interactions.  The model we are thus led to is one of the
earliest (and conceptually simplest) preonic models proposed
[9]: a fermion is assumed to be a composite of three types
of spin $\frac{1}{2}$ preons, $|i,f,\alpha \rangle = | i \rangle
\otimes |f \rangle \otimes | \alpha \rangle$.  All of them are
subject to a meta-interaction which is responsible for the
binding (and about which it is futile to speculate at this
stage).  In addition, the flavour preons $|f \rangle$ have
an $SU(2)$ gauge interaction and the quark-like colour
preons $|\alpha =c\rangle ,~c = 1,2,3$ have an $SU(3)$
gauge interaction.  If the preons are all $L$-chiral and
the binding is assumed to be without ``orbital excitation'',
the composites necessarily belong to one of the irreducibles
in the representation
$$\left ( \frac{1}{2} , 0 \right ) \otimes \left (
\frac{1}{2} , 0 \right ) \otimes \left ( \frac{1}{2} , 0
\right ) = \left ( \frac{3}{2} , 0 \right ) \oplus 
\left ( \frac{1}{2} , 0 \right ) \oplus \left ( \frac{1}{2}
, 0 \right ) $$
of the Lorentz group.  Since there are good reasons why
massless particles of spin $\frac{3}{2}$ are unlikely to exist
[17], we end up with the right number of $L$-chiral spin
$\frac{1}{2}$ fermions if we assume that only one of the two
$(\frac{1}{2},0)$
representations binds.  In any case, chiral symmetry is valid
strongly and the composite fermions are strictly massless.
We have then a ``metaflavour'' group $U(8N)$ which, when the
effects of instantons are taken into account, leaves $SU
(8N)$ as the group of exact symmetries of all interactions
of the composite fermions [10].  Thus a preonic picture provides
a natural reason why the unifying group $\cg$ is indeed
$SU(8N)$.

It is nevertheless important to highlight the one significant
difference between the picture that emerges here and the
one visualised by 't Hooft [10].  Once it is accepted that
the universality properties of $G$ gauge theory reflect the
existence of flavour and colour preons $|f \rangle$ and
$|\alpha \rangle$ we are obliged to ascribe the (low energy)
$G$-gauge interactions to the preons themselves.  They are
not some residual interactions left over from the meta-interactions
binding them, even less the manifestation of ``spectator''
gauge bosons.  (An instructive down-to-earth parallel is
provided by a set of nuclei having the same mass number but
differing atomic numbers -- their electromagnetic
interactions at energies low enough for nuclear structure
effects to be ignored arise from and are fully determined by
the electromagnetic interactions of the proton and the
neutron).  In particular, the gauge bosons of $G$ are
themselves elementary.

We conclude by noting that there is no conceptual problem in
having a ``grand'' unifying group $\cg$ of which only a
subgroup $G$ is actually a gauge group with gauge bosons
associated to its generators.  The determination of the
gauge coupling constants of $G$ by embedding it in $\cg$ is
independent of whether all of $\cg$ is gauged.  The matrix
elements of universality-violating currents are suppressed,
not because they couple to superheavy gauge bosons, but
because the states between which the matrix elements are
taken respect universality.  Some of the problems occurring
in traditional ``grand'' unification models like the
naturalness and hierarchy problems are then no longer relevant.
Also, we need no longer worry about anomalies connected with
$\cg$ as along as the $G$ theory is anomaly-free, which it
is.  

\noindent
{\bf Acknowledgements.}  This paper has benefited from the
discussions I have had with G.~Rajasekaran.  I thank the
Institute of Mathematical Sciences, Chennai, for the use of
its facilities and the Universities of Hawaii and Oregon for
hospitality at times in the past during which the ideas of
this paper were being worked upon.

\section*{Appendix}

We describe here some elementary group theory used in the
paper.  Let $\cg$ be a group, $S$ a subset of $\cg$, not
necessarily a subgroup.  Then the centraliser of $S$ in
$\cg$ is 
$$C(S) = \{ g \in \cg \mid gs = sg, ~~ \forall ~s \in S \}
.$$
$C(S)$ is a subgroup of $\cg$ containing the centre of
$\cg$.  If $S_1$ and $S_2$ are subsets of $\cg$ and $S_1
\subset S_2$, then obviously, $C(S_2) \subset C(S_1)$.
Consider the centraliser of $C(S)$:
$$C(C(S)) = \{ g\in \cg \mid gx=xg, ~\forall ~x \in C(S)\} .$$
$C(C(S))$ has $S$ as a subset.  Moreover, since every $x
\in C(S)$ commutes with every $g \in C(C(S))$, $C(S)$
centralises $C(C(S))$ and $C(C(S))$ is the maximal subgroup
of $\cg$ containing $S$ and centralised by $C(S)$.  This
observation is useful if we wish to find a subgroup $S$
such that $C(S)$ is a given subgroup, in our case $G_{wk}
\times G_{st} = C(S_{wk}) \times 
C(S_{st})$.  We have $C(C(S)) = \{ g \in \cg \mid gx = xg$,
~$\forall$ $x \in G_{wk}$ {\it and} $gy =yg$, $\forall$ $y
\in G_{st} \} = C(G_{wk}) \cap C(G_{st}) = C(C (S_{wk}))
\cap C(C(S_{st}))$.  It follows that $S = S_{wk} \cap S_{st}$
is the maximal subgroup centralised by $G_{wk} \times
G_{st}$.

Writing the defining representation space of $SU (8N)$
as
$$V = V_{fam} \otimes V_{fl} \otimes (V_l \oplus V_q)$$
we may characterise $S_{wk}$ and $S_{st}$ as
$$S_{wk} = D_{fl}^2 U (V_{fam} \otimes (V_l \oplus
V_q)),$$
$$S_{st} = U(V_{fam} \otimes V_{fl} ) \times D_q^3 (V_{fam}
\otimes V_{fl}).$$
As the notation makes clear, $S_{wk}$ is the diagonal
product with respect to flavour of the unitary group of
$V_{fam} \otimes V_{cal}$ and similarly for $S_{st}$.  The
advantage of this explicit notation is that we can read off
the intersection of $S_{wk}$ and $S_{st}$ as the group
$$S = S_{wk} \cap S_{st} = D^6 U(V_{fam}) \times
D^2U(V_{fam}),$$
where, in the two factors on the right, the diagonal
products are over $V_{fl} \otimes V_q$ and $V_{fl} \otimes
V_l$ respectively.  Therefore $S$ is isomorphic to $U(N)
\times U(N)$.  

In the argument given in section 2 for finding $G$ given
$S_{wk}$ and $S_{st}$, these general considerations were
unnecessary and were dispensed with.  They become very useful
however in finding a minimal Higgs scheme for reducing
$\cg$ to $G$.  For this purpose, we begin by noting the
standard identification of the Lie algebra of $\cg$ with the
vector space of the adjoint representation $\rho_{ad}$ of
$\cg$.  If $\{t_A\}$ is a basis of matrices for Lie $\cg$
and $\xi \in V_{ad}$ is a vector in the adjoint
representation, with components $\{ \xi_A\}$, this
identification is
$$\xi \longleftrightarrow \sum_A \xi_A t_A = X.$$
On the matrix $X$, the adjoint action of $\cg$ corresponds
to the usual one of conjugation by $g \in \cg$:
$$\rho_{ad} (g) \xi \longleftrightarrow g ~X~ g^{-1}.$$
Under this identification, therefore, the centraliser of
$\exp (iX) \in \cg$ corresponds to the set $\{ g \in \cg \mid
\rho_{ad} (g) \xi = \xi \}$, in other words, the little group
of $\xi$.

On the other hand, we have the general characterisation of
the non-zero vacuum values of the Higgs fields as a
numerical vector $\eta$ of the representation (in general
reducible) of $\cg$ to which the Higgs belongs, such that
the little group of $\eta$ in $\cg$ is the unbroken
subgroup $G$.  Hence when $G$ is given as the centraliser of
some subgroup $S$, we may conclude that i) the Higgs fields
can be assigned to (sums of copies of) the adjoint
representation of $\cg$, and ii) their nonvanishing vacuum
values can be chosen to be a set of matrices spanning the Lie
algebra of $S$.  It follows that the Higgs vacuum values which will
break $\cg = SU(8N)$ to $G = SU(2) \times U(1) \times SU(3)$
form two sets of matrices each spanning the Lie algebra of
$SU(N)$ [the centraliser of $SU(N)$ automatically
centralises $U(N)$]; each set consists of $N^2-1$ linearly
independent hermitian $N \times N$ matrices.

\end{document}